\documentclass[aps,prd,nofootinbib,preprintnumbers]{revtex4}
\usepackage{amsmath}
\usepackage{amssymb}
\usepackage{amsfonts}
\usepackage{graphicx}
\usepackage{bm}

\newcommand{\be}{\begin{equation}}
\newcommand{\ee}{\end{equation}}
\newcommand{\bea}{\begin{eqnarray}}
\newcommand{\eea}{\end{eqnarray}}
\newcommand{\beaa}{\begin{eqnarray*}}
\newcommand{\eeaa}{\end{eqnarray*}}

\newcommand{\nn}{\nonumber\\}

\begin{document}


\preprint{YITP-11-72}

\title{Screening of cosmological constant
in non-local cosmology}

\date{\today}
\author{Ying-li Zhang\footnote{E-mail
 address: yingli{}@{}yukawa.kyoto-u.ac.jp}
and
Misao Sasaki\footnote{E-mail
 address: misao{}@{}yukawa.kyoto-u.ac.jp}
}

\affiliation{Yukawa
Institute for Theoretical Physics,
Kyoto University, Kyoto 606-8502, Japan
}

\begin{abstract}
We consider a model of non-local gravity
with a large bare cosmological constant, $\Lambda$,
and study its cosmological solutions.
The model is characterized by a function $f(\psi)=f_0 e^{\alpha\psi}$
where $\psi=\Box^{-1}R$ and $\alpha$ is a real dimensionless parameter.
In the absence of matter, we find an expanding universe
solution $a\propto t^n$ with $n<1$, that is, a universe with decelerated
expansion without any fine-tuning of the parameter.
Thus the effect of the cosmological constant is effectively
shielded in this solution. It has been known that solutions in
non-local gravity often suffer from the existence of ghost modes.
In the present case we find the solution is ghost-free if
$\alpha>\alpha_{cr}\approx0.17$. This is quite a weak condition.
We argue that the solution is stable against
the inclusion of matter fields.
Thus our solution opens up new possibilities for solution
to the cosmological constant problem.
\end{abstract}

\maketitle

\section{introduction}

After the discovery of the accelerated expansion of the current
universe, there have been a lot of discussions about which side
of the Einstein equation should be modified, either by adding
an extremely small cosmological constant or something similar
to the matter side of the equation or by modifying the
gravity side of the equation. In particular, inspired by
string theory which lives in higher dimensions, modified gravity
theories have been attracting much attention in recent years.
One of such is non-local gravity. Its cosmological effect
was studied in~\cite{Deser:2007jk}, and various aspects
of it have been studied~\cite{Non-local-gravity}.

Here we are interested in a different aspect of non-local gravity.
By studying phenomenological modifications of the Einstein
equation, it was suggested that non-local gravity may solve the
cosmological constant problem~\cite{ArkaniHamed:2002fu}.
Inspired by this work, an explicit mechanism to screen the cosmological
constant in non-local gravity was discussed in detail~\cite{Nojiri:2010pw}.
Unfortunately, however, similar to the situation of
many higher derivative theories~\cite{Simon:1990},
non-local gravity often suffers from the existence of a mode with the
incorrect sign of the kinetic term, that is, a ghost.
There has been some attempt to remedy this defect, for example,
by adding an $F(R)$ term~\cite{Nojiri:2011}, but it has
not been successful so far. However, it should be noted that
the analysis was done only for modes around
either flat or de Sitter (constant curvature) spacetime solutions.

In this paper, we consider a simple one-parameter family
of non-local gravity models with cosmological constant $\Lambda$
and study their cosmological solutions.
The models are characterized by a function $f(\psi)=f_0e^{\alpha\psi}$
where $\psi\equiv\Box^{-1}R$, with $\alpha$ being the real,
dimensionless parameter: See Eq.~(\ref{action}) below.

We assume the universe is spatially flat, and consider the
case where there is no matter contribution. Very interestingly,
we find a power-law solution $a\propto t^n$, where $a$ is the
cosmic scale factor, with $n<1$. This implies that the effect of
the cosmological constant is completely shielded to render the
expansion of the universe decelerated. To be specific, as $\alpha$
varies from $-\infty$ to $+\infty$, $n$ increases monotonically
from $0$ to $1/2$. Thus the universe behaves like a radiation-dominated
universe for $\alpha\gg1$. Then we examine if
the solution is free from a ghost. To our happy surprise,
it is found that for a very wide range of the parameter $\alpha$,
namely for $\alpha>\alpha_{cr}\approx0.17$, the solution is
found to be ghost-free. For this range of $\alpha$, we find
$n>n_{cr}\approx0.35$.

The paper is organized as follows.
In Section 2, we present the action for a general class of non-local gravity
without specifying the form of the function $f(\psi)$,
and derive the equations of motion for spatially flat cosmology.
Then we derive the ghost-free condition of the model
by making a conformal transformation from the original (Jordan)
frame to the Einstein frame.
In Section 3, we specialize the theory to a model characterized
by the $f(\psi)=f_0e^{\alpha\psi}$ and derive its cosmological
solutions. We then examine the ghost-free condition.
Section 4 is denoted to conclusion.

\section{non-local gravity and ghost-free condition}
\subsection{Action and equations of motion}

We consider a class of non-local gravity whose action is given by
\be \label{action}
S=\int d^4 x \sqrt{-g}\left\{
\frac{1}{2\kappa^2}\left[ R\left(1 + f(\Box^{-1}R)\right)
-2\Lambda \right] + \mathcal{L}_\mathrm{matter} \left(Q; g\right)
\right\}\, ,
\ee
where $\kappa^2=8\pi G$, $f$ is a function that characterizes the
nature of non-locality with $\Box^{-1}$ being the inverse of
the d'Alembertian operator, $\Lambda$ is a (bare) cosmological constant
and $Q$ stands for matter fields. For definiteness, we assume
matter is coupled minimally to gravity. Therefore, the above
may be regarded as an action in the Jordan frame.

In this simple class of non-local gravity, we may
rewrite the action into a local form by introducing
two scalar fields $\psi$ and $\xi$ as
\bea\label{action2}
S&=&\int d^4 x \sqrt{-g}\left[
\frac{1}{2\kappa^2}\left\{R\left(1 + f(\psi)\right) -
\xi\left(\Box\psi - R\right)- 2 \Lambda
\right\} + \mathcal{L}_\mathrm{matter}  \right] \nn
&=& \int d^4 \sqrt{-g}\left[ \frac{1}{2\kappa^2}\left\{R\left(1 +
f(\psi)+\xi\right) + g^{\mu\nu}\partial_\mu \xi \partial_\nu \psi
- 2 \Lambda \right\}+ \mathcal{L}_\mathrm{matter} \right] \, .
\eea

By varying the action with respect to $g_{\mu\nu}$, $\xi$ and
$\psi$, respectively, one obtains the field equations as
\bea
0 &=& \frac{1}{2}g_{\mu\nu} \left\{R\left(1 +
f(\psi)+ \xi\right)
 + g^{\alpha\beta}(\partial_\alpha \xi \partial_\beta \psi) - 2 \Lambda
\right\}- R_{\mu\nu}\left(1 + f(\psi)+\xi\right) \nn
&& - \frac{1}{2}\left(\partial_\mu \xi \partial_\nu \psi +
\partial_\mu \psi \partial_\nu \xi\right)
 -\left(g_{\mu\nu}\Box - \nabla_\mu \nabla_\nu\right)\left( f(\psi)
 +\xi\right)+ \kappa^2T_{\mu\nu}\, ,
\label{nl4}
\\
0&=&\Box\psi-R\,,
\\
0&=&\Box\xi- f'(\psi) R\, ,
\label{nl5}
\eea
where $f'(\psi)\equiv df/d\psi$.

For this class of models, we consider their cosmological solutions.
We assume a spatially flat FLRW universe with the metric,
\begin{eqnarray}
ds^2=-dt^2+a^2(t)\delta_{ij}dx^idx^j\,.
\end{eqnarray}
With this assumption, Eqs.~(\ref{nl4})-(\ref{nl5}) reduce to
\bea
\label{einstein1} 0 &=& - 3 H^2\left(1 + f(\psi) + \xi\right) -
\frac{1}{2}\dot\xi \dot\psi
 - 3H\left(f'(\psi)\dot\psi + \dot\xi\right) + \Lambda
+ \kappa^2 \rho\, ,\\
\label{einstein2} 0 &=& \left(2\dot H + 3H^2\right) \left(1 +
f(\psi) + \xi\right) - \frac{1}{2}\dot\xi \dot\psi +
\left(\frac{d^2}{dt^2} + 2H \frac{d}{dt} \right) \left( f(\psi) +\xi
\right)
 - \Lambda + \kappa^2 P\,,\\
\label{psieq}0 &=& \ddot \psi + 3H \dot \psi + 6 \dot H + 12 H^2 \, , \\
\label{xieq} 0 &=& \ddot \xi + 3H \dot \xi + \left( 6 \dot H + 12
H^2\right)f'(\psi) \,,
\eea
where a dot denotes the time derivative $\dot{}=d/dt$,
$H=\dot a/a$, and $\rho=-T^0_0$ and $P=T^i_i/3$ are the
 energy density and pressure of the matter fields, respectively.

\subsection{Ghost-free condition}

One may be able to find various interesting solutions
of the theory given by the action~(\ref{action})
or (\ref{action2}),
they may not be physically relevant if they contain
a ghost mode, that is, a mode with a wrong sign of
the kinetic term. Such a mode may be harmless at classical
level, but it would lead to a disastrous consequence
in general as soon as we quantize it.
Hence it is better to avoid the existence of a ghost.
The ghost-free condition for the theory~(\ref{action2})
was discussed in~\cite{Nojiri:2010pw}. Here for completeness,
let us briefly summarize the result.

To examine if the theory contains a ghost or not, it is
convenient to make a conformal transformation of the metric
to bring the action into the one in the Einstein frame, namely
the conformal frame in which the gravitational part of
the action (\ref{action2}) becomes purely Einstein.

For this purpose, in this subsection (and only in this subsection),
we denote the metric in the Jordan frame by $\tilde{g}_{\mu\nu}$,
and consider a conformal transformation,
\begin{eqnarray}
\tilde g_{\mu\nu}=\Omega^2 g_{\mu\nu}\,.
\label{tildeg}
\end{eqnarray}
Then we have
\bea\label{conformtrans}
\tilde{R}= \Omega^{-2}\left[R-6(\Box\log{\Omega}
+g^{\mu\nu}\nabla_\mu\log{\Omega}\nabla_\nu\log{\Omega})
\right]\,.
\eea
Therefore we find that if we identify the conformal factor as
\bea
\Omega^{-2}=1+f(\psi)+\xi\,,
\label{Omega}
\eea
the gravitational part of the action becomes Einsteinian.

The action in the Einstein frame is given by
\bea
S=\int d^4x\sqrt{-g}\left\{\frac{1}{2\kappa^2}\left[R-6(\Box\log{\Omega}+
g^{\mu\nu}\nabla_\mu\log{\Omega}\nabla_\nu\log{\Omega})+
\Omega^2g^{\mu\nu}\nabla_\mu\xi\nabla_\nu\psi-2\Omega^4\Lambda\right]
+\Omega^4\mathcal{L}_\mathrm{matter}(Q; \Omega^2g)\right\}\,.
\eea
Introducing a new field $\phi$ by
\bea\label{conformfactor}
\phi\equiv\log{\Omega}=-\frac{1}{2}\log{\left(1+f(\psi)+\xi\right)}\,,
\eea
in place of $\xi$, the above may be written as
\bea\label{finalver}
S=\int d^4x\sqrt{-g}\left\{\frac{1}{2\kappa^2}\left(R-6
\nabla^\mu\phi\nabla_\mu\phi-2\nabla^\mu\phi\nabla_\mu\psi
-e^{2\phi}f'(\psi)\nabla^\mu\psi\nabla_\mu\psi-2e^{4\phi}\Lambda\right)
+e^{4\phi}\mathcal{L}_\mathrm{matter}(Q; e^{2\phi}g)\right\}\,.
\eea

It is now easy to derive the ghost-free condition.
Since there are only two scalar fields, the condition for
the absence of a ghost is that the trace and the determinant
of the kinetic term matrix are both positive.
In the present case, it is readily seen that
only the positivity of the determinant is sufficient,
that is,
\begin{eqnarray}\label{det}
\det\left|
  \begin{array}{ccc}
    6 & 1 \\
    1 & e^{2\phi}f'(\psi) \\
  \end{array}
\right|>0\,.
\end{eqnarray}
In terms of the original fields, this condition
is expressed as
\bea\label{noghost}
f'(\psi)>\frac{1}{6}(1+f(\psi)+\xi)>0\,,
\eea
where $1+f(\psi)+\xi>0$ is a necessary condition
from Eq.~(\ref{conformfactor}).
Later the above condition is used to examine if
our cosmological solutions are free from ghosts or not.

\section{cosmological solutions}

As a class of simple non-local gravity models, we consider the case
when $f(\psi)$ is given by an exponential function,
\begin{eqnarray}
f(\psi)=f_0e^{\alpha\psi}\,.
\label{expmodel}
\end{eqnarray}
For this class of models, we look for cosmological solutions
with the assumption that the scale factor is a power-law
function of time $t$, $a\propto t^n$. We also assume the
absence of matter fields, $\rho=P=0$.

\subsection{cosmological solutions}

\begin{figure}
\includegraphics[height=7cm ,keepaspectratio=true,angle=0]{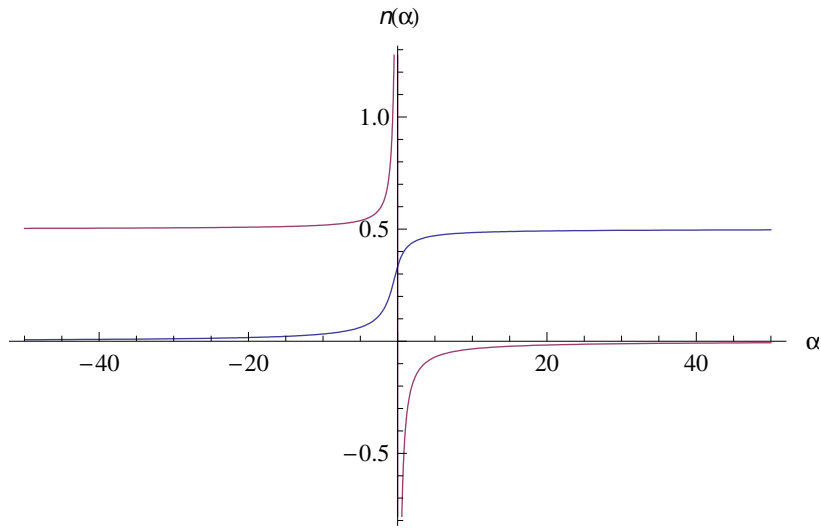}
\caption{Index $n$ as a function of parameter $\alpha$.
Here the blue line (the one in the middle, which is regular
at $\alpha=0$) denotes $n=n_1$ and the red line $n=n_2$.}
\label{index}
\end{figure}

Under the assumption that $a\propto t^n$, we solve the system of
equations (\ref{einstein1}) - (\ref{xieq}).
First we solve Eq.~(\ref{psieq}). Inserting $H=n/t$ to it,
we find the solution for $\psi(t)$ as
\bea\label{psisol}
\psi(t)&=&\psi_1t^{1-3n}-\frac{6n(2n-1)}{3n-1}\ln{t/t_0}\,,
\eea
where $\psi_1$ and $t_0$ are two integral constants.
For simplicity, we set $\psi_1=0$.
It may be worth mentioning that $\psi$ is the variable that signifies
the nonlocality of our theory, $\psi=\Box^{-1}R$. The above choice
of the integration constant corresponds to the (regularized)
retarded integral of the inverse d'Alembertian operator.
Then the function $f(\psi(t))$ takes the form,
\bea
f=f_0\left(\frac{t}{t_0}\right)^{m}\,;
\quad
m\equiv-\alpha \frac{6 n(2n-1)}{3n-1}\,.
\eea
Inserting the above into Eq.~(\ref{xieq}),
 one obtains the solution for $\xi(t)$
\bea
\xi(t)=\xi_0+\xi_1t^{1-3n}
-\frac{6n(2n-1)}{m(3n+m-1)}\alpha f_0\left(\frac{t}{t_0}\right)^{m}\,,
\eea
where $\xi_0$ and $\xi_1$ are two integral constants.
Again for simplicity, we set $\xi_1=0$ in the following.
Inserting these solutions into Eq.~(\ref{einstein1}),
we find for consistency,
\bea\label{integralconst}
\xi_0=-1\,,\quad m=2\,,\quad t_0^2=\frac{6n(n+1)}{\Lambda}f_0\,.
\eea
 From the viewpoint of Eq.~(\ref{xieq}), the above solution
corresponds to the retarded integral of the inverse d'Alembertian operator
plus a specific value of the constant homogeneous solution $\xi_0=-1$.
In particular the second equation determines
the power-law index $n$,
\bea\label{eqn}
6\alpha n^2+3(1-\alpha)n-1=0\,.
\eea
This equation has two real solutions, which are given by
\bea
n_1&=&\frac{-3+3\alpha+\sqrt{3(3\alpha^2+2\alpha+3)}}{12\alpha}\,,
\nonumber\\
n_2&=&\frac{-3+3\alpha-\sqrt{3(3\alpha^2+2\alpha+3)}}{12\alpha}\,.
\eea We plot them in Fig.~\ref{index} as a function of $\alpha$.
From this figure, we see that an expanding solution is given by
$n=n_1$ for any $\alpha$ and by $n=n_2$ for $\alpha<0$. In both
cases, $n$ approaches $1/2$ for $|\alpha|\gg1$, that is, the
evolution of the universe looks like a radiation-dominated one.
We note that from Eqs.~(\ref{integralconst}) and (\ref{eqn}),
the solution for $\psi$ can be rewritten in the form,
\bea
\psi(t)=
\frac{1}{\alpha}\ln\left[\frac{\Lambda t^2}{6f_0n(1+n)}\right]\,.
\eea

Now let us examine the ghost-free condition~(\ref{noghost}).
Noting that $1+f+\xi$ is given by
\begin{eqnarray}
1+f+\xi=f\left(1-\alpha\frac{3n(2n-1)}{3n+1}\right)
=f\frac{6n}{3n+1}\,,
\end{eqnarray}
we have
\bea
\frac{6f'}{1+f+\xi}=\alpha\frac{3n+1}{n}
=\alpha(3+\frac{1}{n})\,.
\eea
Then an analytical calculation shows that
\bea\left.\frac{6f'}{1+f+\xi}\right|_{n_1}&>&1
~\Longrightarrow~
\alpha>\alpha_{cr}\equiv
\frac{1}{15}(7-2^{\frac{1}{3}}-2^{\frac{5}{3}})\approx0.17\,,
\\
\left.\frac{6f'}{1+f+\xi}\right|_{n_2}&<&1\,
\quad\mbox{for}~\forall\alpha\in\mathbf{R}\,.
\eea
So the ghost-free condition~(\ref{noghost}) implies
that a ghost can be avoided for $n=n_1$ provided that
the parameter $\alpha>\alpha_{cr}\approx0.17$,
while the appearance of a ghost cannot be avoided at all for $n=n_2$.

In summary, we conclude that in our non-local gravity model
with $f(\psi)=f_0e^{\alpha\psi}$, we have a cosmological solution,
\begin{eqnarray}
 \psi(t)=\frac{1}{\alpha}\ln{\left[\frac{\Lambda t^2}{6f_0n(n+1)}\right]}\,,
\quad
 \xi(t)=-1-\frac{(2n-1)\alpha\Lambda t^2}{2(3n+1)(n+1)}\,,
\label{eq:1}
\end{eqnarray}
where
\begin{eqnarray}
 n=n(\alpha)\equiv
\frac{-3+3\alpha+\sqrt{3(3\alpha^2+2\alpha+3)}}{12\alpha}\,,
\end{eqnarray}
and the absence of a ghost is guaranteed provided that the parameter
$\alpha$ satisfies
\begin{eqnarray}
\alpha>\alpha_{cr}=\frac{1}{15}(7-2^{\frac{1}{3}}-2^{\frac{5}{3}})
\approx0.17\,.
\end{eqnarray}
Correspondingly, the power-law index $n$ varies in the range,
\begin{eqnarray}
1/2>n>n_{cr}\approx0.35\,.
\end{eqnarray}
It should be noted that Eq.~(\ref{noghost}) implies $f'(\psi)>0$,
which reduces to the constraint on the parameter $f_0>0$
for a positive $\alpha$. Thus the solution we have
obtained in the above is meaningful only for
a positive cosmological constant $\Lambda$.

Finally, let us comment on the case when a matter field is present.
If the coupling is as assumed in Eq.~(\ref{action}), that is,
if matter is minimally coupled to gravity in the original conformal
frame, then assuming the equation of state parameter $w=P/\rho$
is greater than minus one, $w>-1$, its effect will be always subdominant
if it is so initially simply because its contribution to the energy
density decreases faster than that of the cosmological constant.
So our solution will be stable against the inclusion of matter fields.

\subsection{behavior in the Einstein frame}

In this subsection, we study the behavior of the
solution~(\ref{eq:1}) in the Einstein frame. The reason
for doing this is the following. Although matter fields are
assumed to be minimally coupled to gravity in the original
frame as given in the action~(\ref{action}), this assumption
has no relevance to our cosmological solution because we simply
neglected their presence. In other words, we still have
freedom in the choice of matter-gravity coupling. In particular,
one may assume that a certain matter field is minimally coupled
to gravity in the Einstein frame. 
Another reason is to confirm explicitly that the expansion is
delecerated in the Einstein frame as well.\footnote{In fact, whether
the expansion is delecerate or not does not depend
on the choice of conformal frames, because it is directly
related to the causal structure of the universe, which is
conformal invariant by definition.}
Therefore it is useful to know
the behavior of our cosmological solution in the Einstein frame.

In this subsection, we attach a subscript $_E$ to quantities in
the Einstein frame.
Under the conformal conformal transformation
$g_{\mu\nu}=\Omega^2g_{E,\mu\nu}$
the conformal factor is given by
\bea\label{confo}
\Omega^2=\frac{1}{1+f+\xi}=\frac{(n+1)(3n+1)}{\Lambda t^2}\,,
\eea
where $t$ is the time in the original frame.

The cosmic proper time $t_E$ in the Einstein frame
is related to $t$ by $dt=\Omega dt_E$.
Hence we obtain
\bea\label{t2trans}
t_E=\int^t \frac{dt}{\Omega}
=\frac{\sqrt{\Lambda}\,t^2}{2\sqrt{(n+1)(3n+1)}}\,,
\eea
where we have identified $t_E=0$ with $t=0$.

The scale factor in the Einstein frame is $a_E=\Omega\, a$.
Hence the Hubble parameter in the Einstein frame can be calculated as
\bea\label{Hein}
H_E=\frac{1}{a_E}\frac{da_E}{dt_E}
=\frac{\Omega}{a}\frac{d}{dt_E}\left(\frac{a}{\Omega}\right)
=\frac{\Omega^2}{a}\frac{d}{dt}\left(\frac{a}{\Omega}\right)
=\frac{n+1}{2}\frac{1}{t_E}\,,
\eea
where we have used the fact $a\propto t^n$ and $\Omega\propto t^{-2}$
together with Eq.~(\ref{t2trans}).
This implies that the power-law index $n_E$ in the Einstein frame is
given by $n_E=(n+1)/2$, and takes the values in the range,
\begin{eqnarray}
0.75>n_E>n_{E,cr}\approx0.675\,,
\end{eqnarray}
for $\infty>\alpha>\alpha_{cr}\approx0.17$.

Thus, from (\ref{t2trans}) and (\ref{Hein}), we obtain
the solutions for the scalar fields $\psi$ in the Einstein frame
\begin{eqnarray}
\psi(t_E)&=&\frac{1}{\alpha}
\ln{\left[\frac{1}{3f_0n}\sqrt{\frac{\Lambda(3n+1)}{n+1}}\,t_E\right]}\,,
\nonumber\\
\xi(t_E)&=&-1-\alpha(2n-1)\sqrt{\frac{\Lambda}{(3n+1)(n+1)}}\,t_E\,.
\label{eq:2}
\end{eqnarray}
Again from Eq.~(\ref{confo}), we note that the cosmological
constant must be positive, $\Lambda>0$.

Before closing this section, an important comment is in order.
As we demonstrated, the theory can be recasted in the form
of a scalar-tensor theory or in the form of Einstein gravity plus
two scalar fields. Therefore one might consider our theory to
be simply a scalar-tensor theory that gives rise to a decaying
cosmological constant. Technically this is true. But it should
be emphasized that the form of the action for these scalar fields
in either the original frame or the Einstein frame is completely fixed
by the original form of non-local gravity given by Eq.~(\ref{action}):
There is no freedom in maneuvering the form of the Lagrangian for these
scalar fields. In other words, even if one were to regard this theory
as a scalar-tenor theory with two non-minimally coupled scalar fields
(in the original Jordan frame), it would be highly non-trivial to find
a theory that would lead to a solution with a decaying cosmological
constant which decays sufficiently fast in the original frame as well as
in the Einstein frame.

\section{conclusion and discussion}
We have studied cosmological solutions in a simple class of
non-local gravity with cosmological constant. The model is
characterized by a function $f(\psi)=f_0e^{\alpha\psi}$ where
$f_0>0$ and $\alpha$ is a real parameter, and $\psi$ is the inverse
of the d'Alembertian acting on the scalar curvature,
$\psi=\Box^{-1}R$. In the absence of matter fields, we have found
power-law solutions $a\propto t^n$ with $n<1$, that is, with
decelerated expansion. We have found that for
$\alpha>\alpha_{cr}\approx0.17$, the solution is ghost-free. Thus
without any fine-tuning, the solution successfully screens the
effect of the cosmological constant that would have led to
accelerated expansion.

We have assumed the absence of matter. If there is matter,
we have to specify how the matter couples to gravity.
If the matter couples minimally to gravity in the original
(Jordan) frame, then it is expected that the presence
of matter will not change the qualitative behavior of
the solution in the sense that the solution we found will
be an asymptotic, attractor solution of the system.

Our solution opens up new possibilities for the solution
to the cosmological constant problem.
It is known that the expansion of the present universe
is accelerating. Therefore, in order to make our model more
realistic, we must have matter which are not minimally coupled
in the original frame, and which would lead to accelerated
expansion at late times (but with a cosmological
constant which is exponentially smaller than the original
cosmological constant). Search for such a model is under way.

\begin{acknowledgments}
This work was supported in part by
the Grant-in-Aid for the Global COE Program
``The Next Generation of Physics, Spun from Universality and Emergence''
from the Ministry of Education, Culture,
Sports, Science and Technology (MEXT) of Japan,
by JSPS Grant-in-Aid for Scientific Research (A) No.~21244033,
and by JSPS Grant-in-Aid for Creative Scientific Research No.~19GS0219.
\end{acknowledgments}


\begin{thebibliography}{99}

\bibitem{Deser:2007jk}
  S.~Deser and R.~P.~Woodard,
  Phys.\ Rev.\ Lett.\  {\bf 99}, 111301 (2007)
  [arXiv:0706.2151 [astro-ph]].

\bibitem{Non-local-gravity}
%
  L.~Joukovskaya,
  Phys.\ Rev.\  D {\bf 76}, 105007 (2007)
  [arXiv:0707.1545 [hep-th]]; \\
%
  S.~Nojiri and S.~D.~Odintsov,
  Phys.\ Lett.\  B {\bf 659}, 821 (2008)
  [arXiv:0708.0924 [hep-th]]; \\
%
  G.~Calcagni, M.~Montobbio and G.~Nardelli,
  Phys.\ Lett.\  B {\bf 662}, 285 (2008)
  [arXiv:0712.2237 [hep-th]]; \\
%
  T.~Koivisto,
  Phys.\ Rev.\  D {\bf 77}, 123513 (2008)
  [arXiv:0803.3399 [gr-qc]]; \\
%
  T.~S.~Koivisto,
  Phys.\ Rev.\  D {\bf 78}, 123505 (2008)
  [arXiv:0807.3778 [gr-qc]]; \\
%
  T.~Biswas, T.~Koivisto and A.~Mazumdar,
  JCAP {\bf 1011}, 008 (2010)
  [arXiv:1005.0590 [hep-th]]; \\
%
  S.~Capozziello, E.~Elizalde, S.~Nojiri and S.~D.~Odintsov,
  Phys.\ Lett.\  B {\bf 671}, 193 (2009)
  [arXiv:0809.1535 [hep-th]]; \\
%
  N.~A.~Koshelev,
  Grav.\ Cosmol.\  {\bf 15}, 220 (2009)
  [arXiv:0809.4927 [gr-qc]]; \\
%
  C.~Deffayet and R.~P.~Woodard,
  JCAP {\bf 0908}, 023 (2009)
  [arXiv:0904.0961 [gr-qc]]; \\
%
  S.~Jhingan, S.~Nojiri, S.~D.~Odintsov, M.~Sami, I.~Thongkool and S.~Zerbini,
  Phys.\ Lett.\  B {\bf 663}, 424 (2008)
  [arXiv:0803.2613 [hep-th]]; \\
%
  G.~Cognola, E.~Elizalde, S.~Nojiri, S.~D.~Odintsov and S.~Zerbini,
  Eur.\ Phys.\ J.\  C {\bf 64}, 483 (2009)
  [arXiv:0905.0543 [gr-qc]]; \\
%
  H.~Farajollashi and F.~Milani,
  Int.\ J.\ Theor.\ Phys.\ {\bf 50}, 1953 (2011)
  [arXiv:1103.3553 [gr-qc]]; \\
%
  J.~Kluson,
  arXiv:hep-th/1105.6056; \\
%
  D.~J.~Liu, B.~Yang and X.~H.~Jin,
  arXiv:gr-qc/1108.2894.


\bibitem{ArkaniHamed:2002fu}
  N.~Arkani-Hamed, S.~Dimopoulos, G.~Dvali and G.~Gabadadze,
  arXiv:hep-th/0209227.

\bibitem{Nojiri:2010pw}
  S.~Nojiri, S.~D.~Odintsov, M.~Sasaki and Y.~l.~Zhang,
  Phys.\ Lett.\  B {\bf 696}, 278 (2011)
  [arXiv:1010.5375 [gr-qc]].

\bibitem{Simon:1990}
  J.~Z.~Simon,
  Phys.\ Rev.\  D {\bf 41}, 3720 (1990).

\bibitem{Nojiri:2011}
  K.~Bamba, S.~Nojiri, S.~D.~Odintsov and M.~Sasaki,
  arXiv:1104.2692 [gr-qc].


\end{thebibliography}
\end{document}